\documentstyle[10pt,epsfig,dp_delphititle,amssymb,subfig]{dp_delphi}
\makeindex  

\def\DpPaperGroup{PH-EP}
\def\DpPaperRef{2006-015}
\def\DpDate{10 May 2006}
\def\DpAuthors{DELPHI Collaboration}
\def\DpSubmit{(Accepted by Phys. Lett. B)}
\def\DpTitle{{ Study of Leading Hadrons in Gluon and Quark Fragmentation

}}
\def\DpComment{}
\def\DpEMail{}

\begin{document}
%%%%%%%%%%%%%%%%%%%%%%%%%% They are a problem with Coll.Sty ?
\makeatletter
\makeatother
%%%%%%%%%%%%%%%%%%%%%%%%%% ??????????????????????????????????
%   Generate the title page

\begin{titlepage}
\pagenumbering{roman}

\CERNpreprint{\DpPaperGroup}{\DpPaperRef}   % Reference of the paper
\date{{\small\DpDate}}              % Date of the paper
\title{\DpTitle}                % Title of the paper
\address{\DpAuthors}                % General name of the author(s)

\begin{shortabs}                % Start the abstract
\noindent
The study of quark jets in $e^+e^-$ reactions at LEP has
demonstrated that the hadronisation process is reproduced well by
the Lund string model. However, our understanding of gluon
fragmentation is less complete. In this study enriched quark and
gluon jet samples of different purities are selected in three-jet
events from hadronic decays of the Z collected by the DELPHI experiment in
the LEP runs during 1994 and 1995. The leading systems of the two kinds
of jets are defined by requiring a rapidity gap and their  sum of
charges is studied. An excess of leading systems with total charge
zero is found for gluon jets in all cases, when compared to Monte
Carlo Simulations with JETSET (with and without Bose-Einstein
correlations included) and ARIADNE. The corresponding leading
systems of quark jets do  not exhibit such an excess. The
influence of the gap size and of the gluon purity on the effect is
studied and a concentration of the excess of neutral leading systems 
at low invariant masses ($\lesssim$ 2 GeV/c$^2$) is
observed, indicating that gluon jets might have an additional
hitherto undetected fragmentation mode via a two-gluon system.
This could be an indication of a possible production of gluonic
states as predicted by QCD.
\end{shortabs}

\vfill

\begin{center}
\DpSubmit \ \\      % Horrible hack to allow to have DpSubmit empty
\DpComment \ \\
\DpEMail \ \\
\end{center}

\vfill
\clearpage

\headsep 10.0pt

\addtolength{\textheight}{10mm}
\addtolength{\footskip}{-5mm}
\begingroup
%           Commands to process the author names
%
\newcommand{\DpName}[2]{\hbox{#1$^{\ref{#2}}$},\hfill}
\newcommand{\DpNameTwo}[3]{\hbox{#1$^{\ref{#2},\ref{#3}}$},\hfill}
\newcommand{\DpNameThree}[4]{\hbox{#1$^{\ref{#2},\ref{#3},\ref{#4}}$},\hfill}
\newskip\Bigfill \Bigfill = 0pt plus 1000fill
\newcommand{\DpNameLast}[2]{\hbox{#1$^{\ref{#2}}$}\hspace{\Bigfill}}

%
%\small
\footnotesize
\noindent
\DpName{J.Abdallah}{LPNHE}
\DpName{P.Abreu}{LIP}
\DpName{W.Adam}{VIENNA}
\DpName{P.Adzic}{DEMOKRITOS}
\DpName{T.Albrecht}{KARLSRUHE}
\DpName{R.Alemany-Fernandez}{CERN}
\DpName{T.Allmendinger}{KARLSRUHE}
\DpName{P.P.Allport}{LIVERPOOL}
\DpName{U.Amaldi}{MILANO2}
\DpName{N.Amapane}{TORINO}
% ...
\normalsize
\endgroup
\newpage

\titlefoot{Department of Physics and Astronomy, Iowa State
     University, Ames IA 50011-3160, USA
    \label{AMES}}
\titlefoot{IIHE, ULB-VUB,
     Pleinlaan 2, B-1050 Brussels, Belgium
    \label{BRUSSELS}}
\titlefoot{Physics Laboratory, University of Athens, Solonos Str.
     104, GR-10680 Athens, Greece
    \label{ATHENS}}
\titlefoot{Department of Physics, University of Bergen,
     All\'egaten 55, NO-5007 Bergen, Norway
    \label{BERGEN}}
\titlefoot{Dipartimento di Fisica, Universit\`a di Bologna and INFN,
     Via Irnerio 46, IT-40126 Bologna, Italy
    \label{BOLOGNA}}
\titlefoot{Centro Brasileiro de Pesquisas F\'{\i}sicas, rua Xavier Sigaud 150,
     BR-22290 Rio de Janeiro, Brazil
    \label{BRASIL-CBPF}}
\titlefoot{Inst. de F\'{\i}sica, Univ. Estadual do Rio de Janeiro,
     rua S\~{a}o Francisco Xavier 524, Rio de Janeiro, Brazil
    \label{BRASIL-IFUERJ}}
\titlefoot{Coll\`ege de France, Lab. de Physique Corpusculaire, IN2P3-CNRS,
     FR-75231 Paris Cedex 05, France
    \label{CDF}}
\titlefoot{CERN, CH-1211 Geneva 23, Switzerland
    \label{CERN}}
\titlefoot{Institut de Recherches Subatomiques, IN2P3 - CNRS/ULP - BP20,
     FR-67037 Strasbourg Cedex, France
    \label{CRN}}
\titlefoot{Now at DESY-Zeuthen, Platanenallee 6, D-15735 Zeuthen, Germany
    \label{DESY}}
\titlefoot{Institute of Nuclear Physics, N.C.S.R. Demokritos,
     P.O. Box 60228, GR-15310 Athens, Greece
    \label{DEMOKRITOS}}
\titlefoot{FZU, Inst. of Phys. of the C.A.S. High Energy Physics Division,
     Na Slovance 2, CZ-180 40, Praha 8, Czech Republic
    \label{FZU}}
\titlefoot{Dipartimento di Fisica, Universit\`a di Genova and INFN,
     Via Dodecaneso 33, IT-16146 Genova, Italy
    \label{GENOVA}}
\titlefoot{Institut des Sciences Nucl\'eaires, IN2P3-CNRS, Universit\'e
     de Grenoble 1, FR-38026 Grenoble Cedex, France
    \label{GRENOBLE}}
\titlefoot{Helsinki Institute of Physics and Department of Physical Sciences,
     P.O. Box 64, FIN-00014 University of Helsinki, 
     \indent~~Finland
    \label{HELSINKI}}
\titlefoot{Joint Institute for Nuclear Research, Dubna, Head Post
     Office, P.O. Box 79, RU-101 000 Moscow, Russian Federation
    \label{JINR}}
\titlefoot{Institut f\"ur Experimentelle Kernphysik,
     Universit\"at Karlsruhe, Postfach 6980, DE-76128 Karlsruhe,
     Germany
    \label{KARLSRUHE}}
\titlefoot{Institute of Nuclear Physics PAN,Ul. Radzikowskiego 152,
     PL-31142 Krakow, Poland
    \label{KRAKOW1}}
\titlefoot{Faculty of Physics and Nuclear Techniques, University of Mining
     and Metallurgy, PL-30055 Krakow, Poland
    \label{KRAKOW2}}
\titlefoot{Universit\'e de Paris-Sud, Lab. de l'Acc\'el\'erateur
     Lin\'eaire, IN2P3-CNRS, B\^{a}t. 200, FR-91405 Orsay Cedex, France
    \label{LAL}}
\titlefoot{School of Physics and Chemistry, University of Lancaster,
     Lancaster LA1 4YB, UK
    \label{LANCASTER}}
\titlefoot{LIP, IST, FCUL - Av. Elias Garcia, 14-$1^{o}$,
     PT-1000 Lisboa Codex, Portugal
    \label{LIP}}
\titlefoot{Department of Physics, University of Liverpool, P.O.
     Box 147, Liverpool L69 3BX, UK
    \label{LIVERPOOL}}
\titlefoot{Dept. of Physics and Astronomy, Kelvin Building,
     University of Glasgow, Glasgow G12 8QQ
    \label{GLASGOW}}
\titlefoot{LPNHE, IN2P3-CNRS, Univ.~Paris VI et VII, Tour 33 (RdC),
     4 place Jussieu, FR-75252 Paris Cedex 05, France
    \label{LPNHE}}
\titlefoot{Department of Physics, University of Lund,
     S\"olvegatan 14, SE-223 63 Lund, Sweden
    \label{LUND}}
\titlefoot{Universit\'e Claude Bernard de Lyon, IPNL, IN2P3-CNRS,
     FR-69622 Villeurbanne Cedex, France
    \label{LYON}}
\titlefoot{Dipartimento di Fisica, Universit\`a di Milano and INFN-MILANO,
     Via Celoria 16, IT-20133 Milan, Italy
    \label{MILANO}}
\titlefoot{Dipartimento di Fisica, Univ. di Milano-Bicocca and
     INFN-MILANO, Piazza della Scienza 3, IT-20126 Milan, Italy
    \label{MILANO2}}
\titlefoot{IPNP of MFF, Charles Univ., Areal MFF,
     V Holesovickach 2, CZ-180 00, Praha 8, Czech Republic
    \label{NC}}
\titlefoot{NIKHEF, Postbus 41882, NL-1009 DB
     Amsterdam, The Netherlands
    \label{NIKHEF}}
\titlefoot{National Technical University, Physics Department,
     Zografou Campus, GR-15773 Athens, Greece
    \label{NTU-ATHENS}}
\titlefoot{Physics Department, University of Oslo, Blindern,
     NO-0316 Oslo, Norway
    \label{OSLO}}
\titlefoot{Dpto. Fisica, Univ. Oviedo, Avda. Calvo Sotelo
     s/n, ES-33007 Oviedo, Spain
    \label{OVIEDO}}
\titlefoot{Department of Physics, University of Oxford,
     Keble Road, Oxford OX1 3RH, UK
    \label{OXFORD}}
\titlefoot{Dipartimento di Fisica, Universit\`a di Padova and
     INFN, Via Marzolo 8, IT-35131 Padua, Italy
    \label{PADOVA}}
\titlefoot{Rutherford Appleton Laboratory, Chilton, Didcot
     OX11 OQX, UK
    \label{RAL}}
\titlefoot{Dipartimento di Fisica, Universit\`a di Roma II and
     INFN, Tor Vergata, IT-00173 Rome, Italy
    \label{ROMA2}}
\titlefoot{Dipartimento di Fisica, Universit\`a di Roma III and
     INFN, Via della Vasca Navale 84, IT-00146 Rome, Italy
    \label{ROMA3}}
\titlefoot{DAPNIA/Service de Physique des Particules,
     CEA-Saclay, FR-91191 Gif-sur-Yvette Cedex, France
    \label{SACLAY}}
\titlefoot{Instituto de Fisica de Cantabria (CSIC-UC), Avda.
     los Castros s/n, ES-39006 Santander, Spain
    \label{SANTANDER}}
\titlefoot{Inst. for High Energy Physics, Serpukov
     P.O. Box 35, Protvino, (Moscow Region), Russian Federation
    \label{SERPUKHOV}}
\titlefoot{J. Stefan Institute, Jamova 39, SI-1000 Ljubljana, Slovenia
    \label{SLOVENIJA1}}
\titlefoot{Laboratory for Astroparticle Physics,
     University of Nova Gorica, Kostanjeviska 16a, SI-5000 Nova Gorica, Slovenia
    \label{SLOVENIJA2}}
\titlefoot{Department of Physics, University of Ljubljana,
     SI-1000 Ljubljana, Slovenia
    \label{SLOVENIJA3}}
\titlefoot{Fysikum, Stockholm University,
     Box 6730, SE-113 85 Stockholm, Sweden
    \label{STOCKHOLM}}
\titlefoot{Dipartimento di Fisica Sperimentale, Universit\`a di
     Torino and INFN, Via P. Giuria 1, IT-10125 Turin, Italy
    \label{TORINO}}
\titlefoot{INFN,Sezione di Torino and Dipartimento di Fisica Teorica,
     Universit\`a di Torino, Via Giuria 1,
     IT-10125 Turin, Italy
    \label{TORINOTH}}
\titlefoot{Dipartimento di Fisica, Universit\`a di Trieste and
     INFN, Via A. Valerio 2, IT-34127 Trieste, Italy
    \label{TRIESTE}}
\titlefoot{Istituto di Fisica, Universit\`a di Udine and INFN,
     IT-33100 Udine, Italy
    \label{UDINE}}
\titlefoot{Univ. Federal do Rio de Janeiro, C.P. 68528
     Cidade Univ., Ilha do Fund\~ao
     BR-21945-970 Rio de Janeiro, Brazil
    \label{UFRJ}}
\titlefoot{Department of Radiation Sciences, University of
     Uppsala, P.O. Box 535, SE-751 21 Uppsala, Sweden
    \label{UPPSALA}}
\titlefoot{IFIC, Valencia-CSIC, and D.F.A.M.N., U. de Valencia,
     Avda. Dr. Moliner 50, ES-46100 Burjassot (Valencia), Spain
    \label{VALENCIA}}
\titlefoot{Institut f\"ur Hochenergiephysik, \"Osterr. Akad.
     d. Wissensch., Nikolsdorfergasse 18, AT-1050 Vienna, Austria
    \label{VIENNA}}
\titlefoot{Inst. Nuclear Studies and University of Warsaw, Ul.
     Hoza 69, PL-00681 Warsaw, Poland
    \label{WARSZAWA}}
\titlefoot{Fachbereich Physik, University of Wuppertal, Postfach
     100 127, DE-42097 Wuppertal, Germany \\
\noindent
{$^\dagger$~deceased}
    \label{WUPPERTAL}}
\addtolength{\textheight}{-10mm}
\addtolength{\footskip}{5mm}
\clearpage

\headsep 30.0pt
\end{titlepage}

%%%%%%%%%%%%%%%%%%%%%%%%%
%
%   Change for the document body
%%\pagestyle{heading}                   % for page numbering
\pagenumbering{arabic}                  % page numbering in number
\setcounter{footnote}{0}                %
\large
%\linenumbers %%%CD
%\input{document.tex}    % The body of the document.
\section{Introduction}

The study 
of quark jets provides us with remarkable insights into the
mechanism of hadronisation. It gives strong evidence for
chain-like charge ordered particle production in excellent
agreement with string Monte Carlo models like JETSET \cite{jet} .
This is shown e.g. by several contributions
\cite{co1,co2,co3,co4} of the DELPHI experiment at LEP, where the compensation
of quantum numbers, in particular that of charge, 
has been extensively studied.  
Much less
is, however, known about the behaviour of gluon jets. On the
theoretical side, besides the fragmentation via two strings as
implemented in JETSET/PYTHIA and ARIADNE \cite{ari}, 
the direct neutralisation of the
color octet field by another gluon with the creation of a
two-gluon system has been considered by Minkowski and Ochs
\cite{ochs1,ochs2} and also by Spiesberger and Zerwas
\cite{zerw}. Older references exist by Montvay \cite{Mont} and 
Peterson and Walsh
\cite{peters}. Additional references can be found in
\cite{ochs1} where it is also emphasized that an experimental study of the
gluon corner in three-jet events could contribute valuably to the question
of the existence of glueballs, an early expectation of QCD \cite{Fri}.
No quantitative prediction however exists up to now.
This has triggered an experimental investigation by
the DELPHI collaboration on gluon fragmentation in a leading
system defined by a rapidity gap \cite{datong,note}. The
preliminary results revealed that electrically neutral systems of
leading particles in gluon jets occur more often than predicted by
JETSET, in agreement with the expectations of the above theoretical
argumentations, while there was no disagreement observed in quark
jets. This phenomenon, experimentally observed for the first
time, has meanwhile also been seen by ALEPH and OPAL
\cite{aleph,opalg}

The JETSET (ARIADNE) model of
 a $q \bar{q}$g event  stretches a string from the q to the g and on to
the $\bar{q}$. The string fragments for example by the creation of $q
\bar{q}$ pairs, similar to what happens for quark fragmentation (Fig.1a).
Thus the JETSET (and ARIADNE)  model regards gluon fragmentation
as a {\em double color triplet fragmentation} (most clearly sketched in Fig.1
of ref.\cite{ochs1}) and the leading
system can obtain the charge $\pm 1$ or 0 in the limiting
configuration. The process proposed by Minkowski
and Ochs, namely the {\em octet neutralisation} of the gluon field
by another gluon has the signature of an {\em uncharged} leading
system due to the requirement that the sum of charges ($SQ$) of the
decay products of a two-gluon system is zero (Fig.1b). In
\cite{ochs1,ochs2} it is also proposed to enhance the possible
contribution of this process  by selecting events where a leading
particle system is separated from the rest of the low energy
particles by a large rapidity gap, empty of hadrons.
In this situation of a hard isolated gluon the octet field is
expected not to have been distorted by multiple gluon emission and
by related color neutralisation processes of small rapidity ranges
\cite{ochs1}. The price to pay for such a selection is, however, a
strong reduction of the number of events because of the Sudakov
form factor \cite{suda}. A different mechanism - color
reconnection \cite{tor} - can produce similar effects. Two
experiments, however, agree that the present color
reconnection models, as implemented in some versions of Monte
Carlo simulations, can not reproduce quantitatively the 
observed excess of $SQ=0$ systems \cite{aleph,opal}.

The present study aims to consolidate the results of the preceding
analyses \cite{datong,note,aleph,opalg} by studying the dependence
of the excess of neutral leading systems in enriched gluon jets on
the gap size and gluon content, by studying their mass spectra 
and by investigating if there are
possible trivial origins for the observed  effect. This is
especially important, since a significant failure of the string
model to describe gluon jets might generally reveal 
the presence of hitherto undetected processes. The size of
the effect for a {\em pure} gluon jet is estimated. As a 
cross-check, the same investigation is done for quark jets. 

\begin{figure}
\mbox{\resizebox{17cm}{!}{\includegraphics{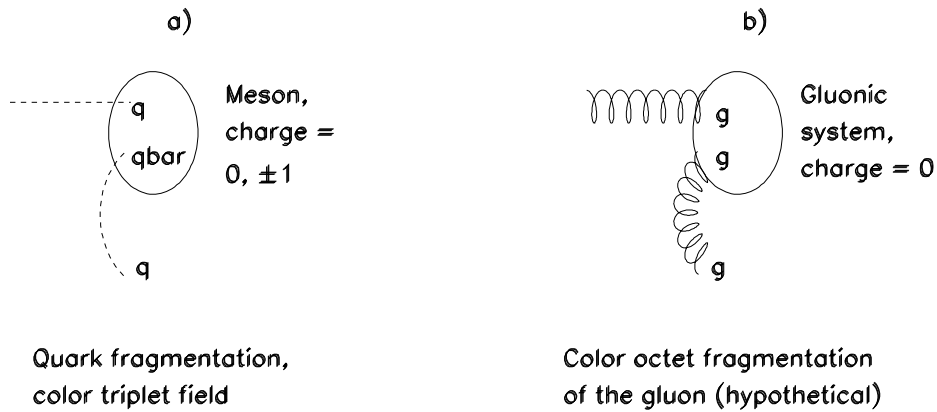}}}
\caption[]{Diagrams to illustrate the processes of color
triplet fragmentation a) and color octet fragmentation 
b). The dashed lines  represent the color triplet
strings and the helixes represent the color octet strings.}
\end{figure}

\section{Data sample and 3-jet event selection}
The data sample used has been collected by the DELPHI experiment at
the LEP collider at the peak of the Z resonance during 1994 and 1995.
Three-jet events have been selected
 by using the appropriate cuts for track quality and for
the hadronic event type \cite{delcut} as well as applying a $k_t$
cluster algorithm (Durham) \cite{durham} 
to all observed charged and neutral particles with $y_{cut}=0.015$ 
\footnote{This value has been obtained from a study optimizing simultaneously
purity and statistics \cite{x1}}.
The
jet energies were recalculated based on the direction of the jet
momenta and the jets were sorted by decreasing energy, i.e. $E_3
\leq E_2 \leq E_1$. Events with $\Theta_2 , \Theta_3  =  135^{\circ} \pm
35^{\circ}$ have been used, where $\Theta_i$  is the inter-jet angle
opposite to jet i. All jets are required to lie in a plane, 
to consist of $\geq 2$ particles and the
jets must be at least $30^{\circ}$ away from the beam direction \cite{x1,x2,x3}.
About 314000 events meet all these conditions.

Without any additional tag the jet with the highest energy $E_1$ (jet1) 
is in most cases a
quark jet and that with the smallest energy $E_3$ (jet3) the gluon
jet. The measured mean jet energies are: $\bar E_1 = 41.4$~GeV,
$\bar E_2 = 32.2$~GeV and $\bar E_3 = 17.7$~GeV.\footnote{Although
the mean energies of jet1 and jet3 differ by more than a factor 2,
the maximum possible rapidities and mean multiplicities 
for charged particles differ
much less (e.g.  $\langle 
nch_{jet3}\rangle$ = 6.04, $\langle
nch_{jet1}\rangle$ =7.36).} In the first data sample ({\bf sample1}), 
where the gluon and quark
jet identification is based on
energy ordering only, events are required not to exhibit any
b-signal (235080 events). Monte Carlo simulations show for the
above mentioned conditions a quark jet contribution of about
90\% 
for jet1\footnote{All quark jet selections (jet1) shown in the figures
for comparisons are defined by sample1.}
 and a gluon jet contribution of about 70\% for jet3.
In a more detailed study of the gluon purity a second independent
sample ({\bf sample2}) is selected, where jet1 and jet2, contrary to jet3, are 
required to
exhibit a b-signal \cite{x3,boris} (Section 4.3). This
additional tag results in a gluon purity of jet3 of about  90\%
and consists of 31400 events.
A third sample ({\bf sample3}) is selected to enable purity unfolding
for special cases. It is defined by the requirement that jet3 has a
b-tag. For this jet3 sample consisting of 12200 events the gluon content 
is very much diminished
(about 26\%). 

\section{Monte Carlo models}
For comparisons a suitable number of Monte Carlo simulations using
JETSET 7.3\cite{jet} and ARIADNE\cite{ari} have been performed. In
contrast to JETSET, ARIADNE incorporates dipole radiation of gluons
instead of the parton shower used by JETSET. Since Bose-Einstein
correlations (BEC) are present in nature, like-charged particles
will stick together in momentum space and local charge
compensation is expected to be diminished. The implementation of
Bose-Einstein correlations into the Monte Carlo simulation,
however, is highly problematic and the magnitude of the effect
 on charge compensation is unknown.
Nevertheless, the possible effect of BEC has to be investigated
and the possible uncertainties have to be considered.

Three different Monte Carlo event samples have been created by
using different generators:

Model (1): JETSET with BEC included (BE32 \cite{be32})
%(about 426000 3-jet events 

Model (2): JETSET without BEC
%(471000)

Model (3): ARIADNE without BEC
%(302000)

The number of events generated for each sample corresponds roughly
to that of the data.

The data are compared to these Monte Carlo event samples with full
simulation of the DELPHI detector. The same reconstruction and
analysis chain has been applied to the data and Monte Carlo (MC)
samples.

\section{Analysis}
\subsection{The sum of charges in the leading system with a 
rapidity gap (sample1)}

After the selection of 3-jet events and the determination of
enriched quark and gluon jet samples, the leading system
of a jet is defined by requiring that all charged 
particles
assigned to the jet
must have a rapidity $y$ with respect to the jet axis 
of $y\geq\Delta y$.
The quantity $\Delta y$ represents a lower limit and defines a
rapidity gap extending at least up to $y = \Delta y$. By this
requirement, also jets are discarded, if they include some particles
(fraction $10^{-3}$) with negative rapidity.
The size of the demanded gap below this leading system 
is a compromise between 
the requirement of a gap as
large as possible and the considerable loss of statistics at a
larger gap. The requirement that the rapidity interval 
$\delta y \ge \Delta y$ (with $\Delta y$ = 1.5) 
below the leading system be empty of charged particles reduces the 
number of jets appreciably.  About 38000 enriched gluon jets and 39000
quark jets meet this condition. 
This reduction rate $f_1$ of enriched gluon jets is quite well 
reproduced by the three Monte Carlo event samples: $f_1$(data) = 0.160,
$f_1$(MC1) = 0.169, $f_1$(MC2) = 0.158, $f_1$(MC3) = 0.157, with the
mean value $f_1$(MC-mean) = 0.161. 
%%The following analysis (with two exceptions in sections 4.4(a)
%%and 5b) is based on
%%charged particles (hadrons and leptons) only.  
In principle, there could be neutral hadrons in the gap.
It has been verified that removing in addition 
topologies where observed neutrals are
contained in the gap
(mainly $\gamma$'s from the decay of $\pi^0$'s),
leads to results that are fully consistent 
with the ones presented here, but with about 15\% larger statistical 
errors.

\begin{figure}
\mbox{\resizebox{17cm}{!}{\includegraphics{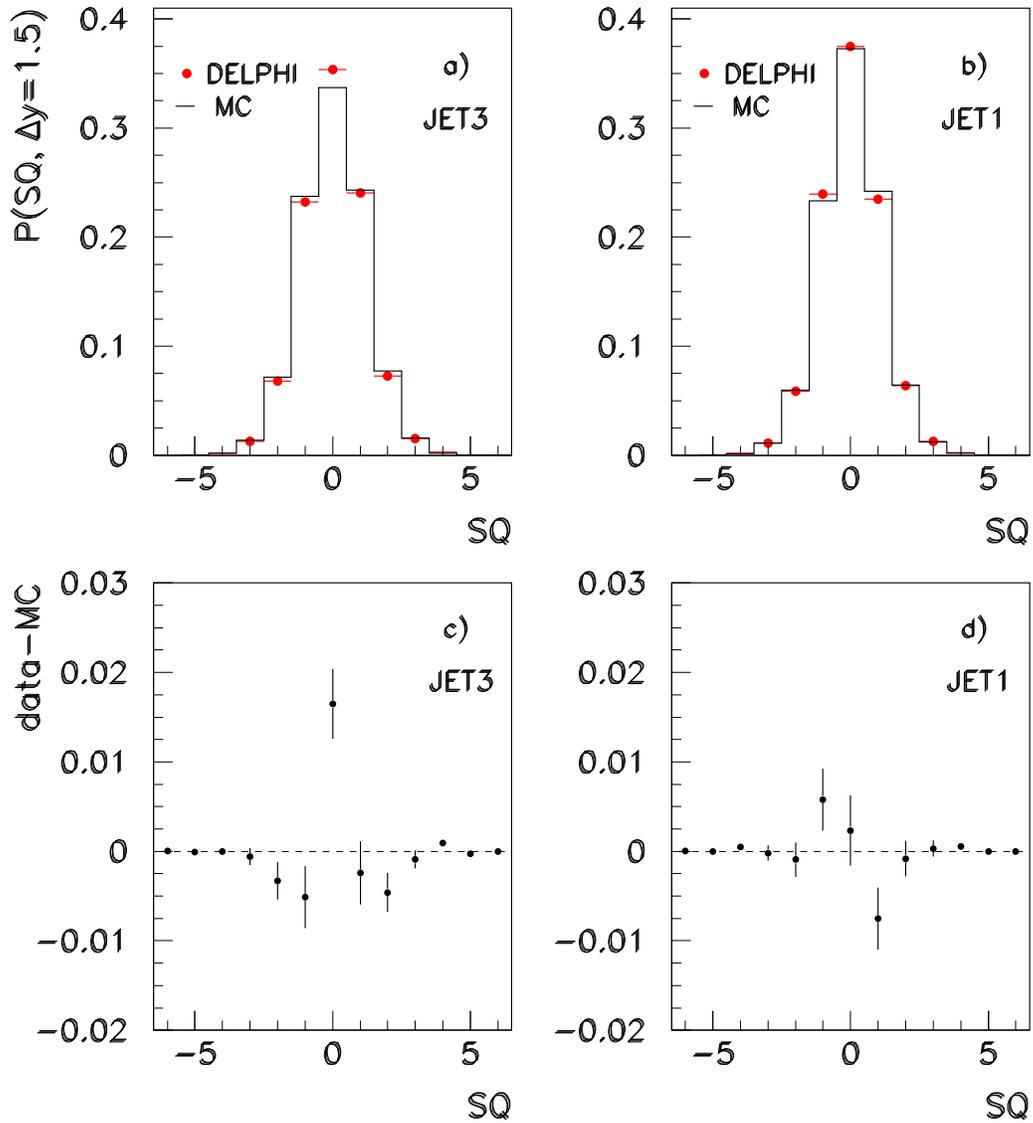}}}
\caption[]{Fraction of jets $P(SQ,1.5)$ as a function of 
the sum of charges $SQ$ of the
leading system for both (a) enriched gluon jets and (b)
quark jets. Full circles represent the data, lines the Monte
Carlo simulation, model (3). 
The  difference
(data-MC) is shown in (c) for enriched gluon jets and in (d) for quark
jets.}
%P(SQ) is defined in Section 4.1.
\end{figure}
The sum of charges $(SQ)$ of the particles belonging to the leading
system defined as above is shown in Fig.2a for enriched gluon jets and in
Fig.2b for quark jets  and compared to ARIADNE.
$P(SQ,\Delta y)$ is generally defined as the
fraction of a jet sample with a gap and a given value of $SQ$, 
\begin{equation}
P(SQ,\Delta y) = \frac{N(SQ,\Delta y)}{N(\Delta y)} 
\end{equation}
and is an                           
estimate for the probability of a jet with a
gap to have a certain $SQ$.
 The $SQ$ distribution  of the leading system for the {\em gluon jet}
(Fig.2a) exhibits for $SQ=0$ a significant enhancement of the data over
the Monte Carlo. This effect is predicted,
%(see Section 1, ref. \cite{ochs1,ochs2}),
if the process of color octet neutralisation is present
\cite{ochs1,ochs2}. On the other hand, there is no significant
difference between the data and the Monte
Carlo simulation in the case of {\em quark jets} (Fig.2b).

The lower parts of Fig.2 show quantitatively the differences of
the $P(SQ,1.5)$ between the data and the Monte Carlo simulation. This
difference for the gluon jet (Fig.2c) amounts to about 4
standard deviations (statistical errors only), for the
quark jet (Fig.2d) this difference is compatible with zero.

\subsection{The dependence on the size of the rapidity gap}

\begin{figure}
\mbox{\resizebox{17cm}{!}{\includegraphics{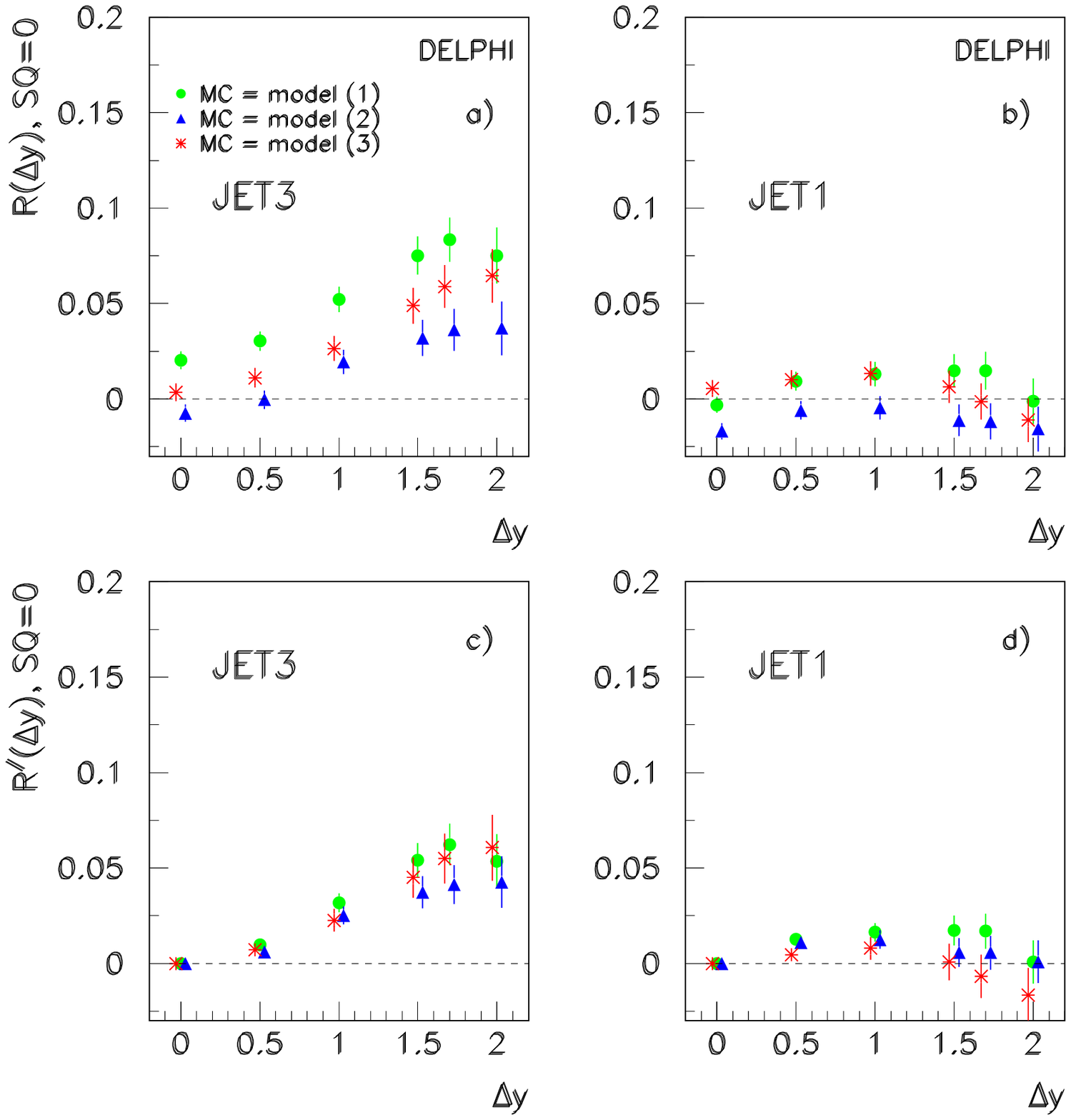}}}
\caption[]{ a),b): Relative deviations $R(\Delta y)$ of the number
of neutral leading systems in gluon and quark jets for the three Monte
Carlo models defined in Section 3 
and for various sizes of $\Delta y$. $R$ is
defined in equ(2). c),d): $R^{'}(\Delta y)$ as defined in
equ(3). Because of the nature of the cut $\Delta y$, the bins are correlated.}
\end{figure}

Figs.3a,b show, for neutral leading systems ($SQ=0$), the dependence
of the relative deviation 
$R(\Delta y)$
on the size of the lower limit ($\Delta y$) of the 
rapidity gaps considered
\footnote{In this representation, the bins are not independent: 
each point represents a subsample of the previous one.}: 

\begin{equation}
R(\Delta y) = \frac{P(0,\Delta y)_{data} - P(0,\Delta y)_{MC}}{P(0,\Delta y)_{MC}} 
\end{equation}
For all three types of Monte Carlo simulations (models (1),
(2) and (3), see Section 3), $R(\Delta y)$ ($\Delta y>0.5$) is
positive and {\em increasing} with $\Delta y$ for jet3 (Fig.3a).
This clearly shows that the surplus of neutral leading
systems in the data, compared to the Monte Carlo simulations, {\em
increases} with the gap size. This corroborates the arguments of
Minkowski and Ochs \cite{ochs1,ochs2}. ARIADNE without BEC (model 3)
lies between the JETSET models. In the case of jet1 (Fig.3b)
all values are scattered around zero and no rise can be seen. When
comparing JETSET with and without BEC included, one
notices for jet3 and also for jet1 a difference 
for all values of $\Delta y$. The
effect of introducing BEC into the Monte Carlo models causes
a shift of $R(\Delta y)$, essentially independent 
of $\Delta y$, to higher values. The
dependence on $\Delta y$ is approximately the same for all three
models. Arguing that only the rise and not implementation effects
of BEC are of interest here and that the surplus of neutral
systems is expected to be small and in a first approximation 
negligible in jets without a gap requirement, the
following quantity is calculated:
\begin{equation}
R^{'}(\Delta y) = R(\Delta y) - R(0)
\end{equation}
and shown in Fig.3c,d.
The residual spread between the models is considered as systematic
uncertainty.

\begin{figure}
\mbox{\resizebox{15.5cm}{!}{\includegraphics{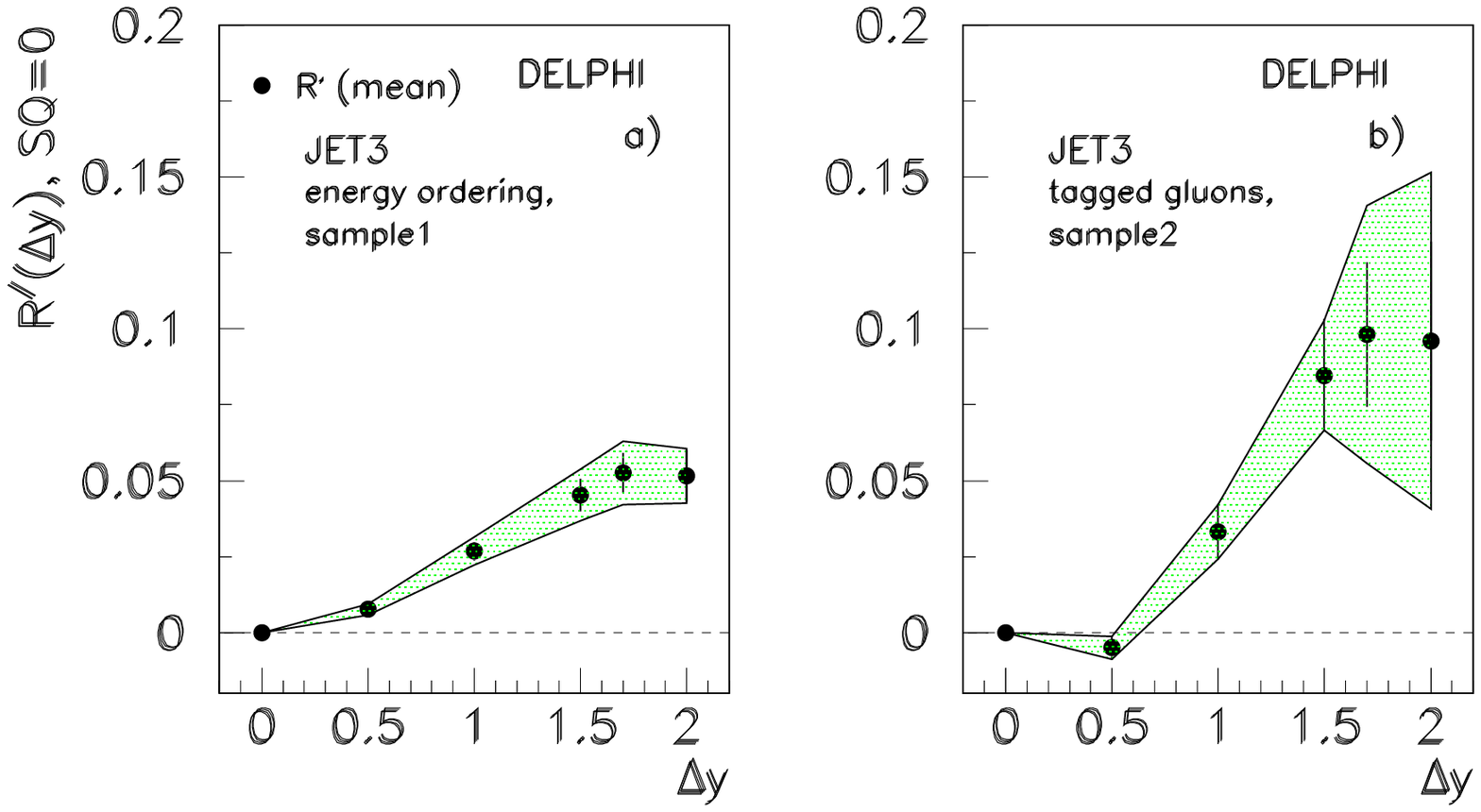}}}
\caption[]{ $R^{'}(\Delta y)$ using  model averages. The gluon jet3 is tagged 
using energy ordering in a) and b-tagging is used for the selection in b). 
}
\end{figure}

\subsection{The dependence of $R^{'}(\Delta y)$ on the gluon purity}
In Fig.4a the mean values of $R^{'}$ with models (1), (2) and (3),
which are presented in Fig.3c for sample1, are drawn together with the
statistical errors (symbols with error bars) and systematic
uncertainty (shaded area). As a cross-check, an independent second
 sample (sample2) of gluon jets with a much higher purity is
selected.
The dependence
of $R^{'}(\Delta y)$ on $\Delta y$ for this
sample is given in Fig.4b.  
Although the statistics are smaller in Fig.4b 
(3870 jet3 at $\Delta y = 1.5$, which is only
about 1/10 of sample1), the effect is
increased, which is expected if it is connected
to the gluon jet only.
At $\Delta y = 1.5$ $R^{'}(\Delta y)$ is
about $0.09 \pm 0.02( statistical)$ because of the higher
purity. 

To estimate the amount of disagreement between data and
Monte Carlo in pure gluon jets the gluon purity has to be
estimated for jet3 at the gap size  $\Delta y$ for
both data selections in Fig.4. In principle, it can be directly
obtained from the Monte Carlo. At the same scale, gluon and quark
jets exhibit different rapidity distributions, i.e. gluon jets
emit more particles per unit at small rapidity. Demanding a gap
reduces therefore not only the number of jets, but also 
the gluon content in a mixed sample of gluon and
quark jets. This is observed in the MC. An estimation of the gluon
purity at gap $\Delta y$ however depends on the correct
modelling of the rapidity distribution of pure gluon jets.
Therefore another method has been used in addition. It uses
the measured reduction rates of the number of jets f$_i$($\Delta y$)
in sample-i by demanding a gap (see Section 4.1) and from the MC only
the composition at gap zero.
Let us define $N_1(\Delta y) = f_1(\Delta y)N_1(0)$ in sample1,  
and $N_2(\Delta y) = f_2(\Delta y)N_2(0)$ in sample2,  
where  $N_1(\Delta y)$  $(N_2(\Delta y))$ is the number of jets
counted at $\Delta y$ in sample1 (sample2). Since these samples are
an admixture of pure q(=light quark) jets, g(=gluon) jets and 
b(=b-quark) jets, the corresponding $f_1(\Delta y),f_2(\Delta y)$ 
(and also $f_3(\Delta y)$ for sample3)
are also an admixture of the reduction rates $f_{q}(\Delta y),
f_{g}(\Delta y), f_{b}(\Delta y)$
of the pure light-quark, pure gluon and pure b-quark subsamples, e.g.
\begin{equation}
f_1(\Delta y)=a_{1q}f_{q}(\Delta y) + a_{1g}f_{g}(\Delta y) +
a_{1b}f_{b}(\Delta y)
\end{equation}
with two analogous equations for $f_2(\Delta y)$ and $f_3(\Delta y)$.
%\footnote{The $f_3(\Delta y)$ has been measured in a third sample,
%where jet3 has a b-tag, this sample however is only used for the
%purity estimation in this section.}
The resulting system of three linear equations can be written in short:
\begin{equation}
F=AF_{pure}
\end{equation}
with the solution
\begin{equation}
F_{pure}=A^{-1}F.
\end{equation}
The vector $F(f_{1}(\Delta y),f_{2}(\Delta y),f_{3}(\Delta y))$ is measured,\\
and the vector $ F_{pure}(f_{q}(\Delta y),f_{g}(\Delta
y),f_{b}(\Delta y))$
is the solution.\\
The matrix A represents the q,g,b compositions for the three
selections at gap=0, estimated from Monte Carlo (e.g. $a_{1g}$ is
the gluon purity of jet3 in sample1, 
 $a_{2g}$ that of sample2, 
and $a_{3g}$ that of sample3
and so on).
With the solution of equ.(6), the numbers of true gluon jets at
$\Delta y$ can be determined in sample1 and sample2: 
\begin{equation}
N_{1}^{gluon}(\Delta y)=f_{g}(\Delta y)a_{1g}N_1(0),
\end{equation}
\begin{equation}
N_{2}^{gluon}(\Delta y)=f_{g}(\Delta y)a_{2g}N_2(0) .
\end{equation}
The fraction of gluon jets $c_{g}^{\Delta y}$ at $\Delta y$ is given by:
\begin{equation}
c_{g}^{\Delta y}(sample1)=N_{1}^{gluon}(\Delta y)/N_1(\Delta y)
=a_{1g}f_{g}(\Delta y)/f_1(\Delta y), 
\end{equation}
\begin{equation}
c_{g}^{\Delta y}(sample2)=N_{2}^{gluon}(\Delta y)/N_2(\Delta y)
=a_{2g}f_{g}(\Delta y)/f_2(\Delta y) .
\end{equation}

Applied to the two data sets in Fig.4a,b the following numbers for
the gluon content have been obtained:
\begin{itemize}
\item[1.] The sample in Fig.4a (sample 1): $c_{g}^{0}=0.65$,
$c_{g}^{1.5}=0.46$ (from equ.(9) ), and $c_{g}^{1.5}=0.45$ (directly from
the Monte Carlo at $\Delta y = 1.5$).
\item[2.] The sample in Fig.4b (sample 2): $c_{g}^{0}=0.88$,
$c_{g}^{1.5}=0.80$ (from equ.(10) ), and $c_{g}^{1.5}=0.82$ (directly from
the Monte Carlo at $\Delta y = 1.5$).
\end{itemize}

\noindent The statistical errors on these numbers are below $1\%$,
systematic errors can be obtained by comparing the estimates with
different Monte Carlo models (1), (2) and (3). They are $\leq 2.6\%$.
The purity estimates obtained above allow the determination of the excess 
of neutral systems $R^{'}_{g}$ in {\em pure} gluon jets
by dividing by $c_{g}^{1.5}$. 
The following values of $R^{'}_{g}$ have been obtained for
the two samples defined above:

\begin{equation}
\mathrm{Sample 1}:\makebox[2cm][r]{\em $R_{g}^{'}$}(1.5) =
0.100\pm0.012 \ (stat) \pm0.025 \ (syst)
\end{equation}
\begin{equation}
\mathrm{Sample 2}:\makebox[2cm][r]{\em $R_{g}^{'}$}(1.5) =
0.107\pm0.022 \ (stat) \pm0.028 \ (syst)
\end{equation}
The samples are statistically independent. Adding statistical and
systematic errors quadratically, a significance of about 3.6$\sigma$ 
is obtained in sample 1 
and of almost 3$\sigma$ in sample 2.
Combining finally both samples results in:
\begin{equation}
\mathrm{Combined}:\makebox[2cm][r]{\em $R_{g}^{'}$}(1.5) =
0.102\pm0.011 \ (stat) \pm0.026 \ (syst)
\end{equation}
This number can be used to make a first estimate of $R_g(0)$,
the amount of the excess of neutral systems in pure gluon jets without
any gap selection.
Taking into account the estimated value of $f_g$(1.5) = 0.112$\pm$0.003
from equ.(6) which tells that 11\% of the pure gluon jets meet
the gap condition $\Delta y$ = 1.5, one obtains:

\begin{equation}
R_g(0) \simeq R_{g}^{'}(\Delta y=1.5)f_g(\Delta y=1.5) = 0.011
\end{equation}

Extending this analysis to samples which allow also for smaller gap sizes 
$1.5 > \Delta y \geq $1.0 leads to $R_g(0)$ values between 0.01 and 0.02.
Extending this furthermore to all gluon jets by taking into account that
$P_g(0,0) \simeq 0.26$ (from sample2, not shown here) leads to the conclusion
that the amount 
of a possible octet neutralisation of the gluon field
is of the order of 0.5\%. 

\subsection{Discussion of the systematic uncertainties}
The following sources of systematic errors have been considered:
\begin{itemize}

\item[(a)]  Quality of event reconstruction.
   Bad reconstructions and losses of particles (mainly neutrals)
   in the detector and wrong
   assignments to the jets can lead to differences
   of several GeV
   between the jet energy calculated from the angles between
   jets ($E_{calc}$) \cite{x3} and the sum of energies of all particles assigned
   to the jet ($E_{sum}$). 
   Improving the quality by cutting away about 1/3 of the jets
   with the largest
   difference $E_{calc}-E_{sum}$ does not significantly  change the signals 
   at $SQ=0$ both in gluon and quark jets.

\item[(b)] The dependence of the effect on the polar angle of the
   jet with respect to the collision axis has been investigated: 
   the effect is stable.

\item[(c)]  The influence of track finding efficiency in the
   detector.
   In order to investigate the influence of track finding efficiency
   the effect of a reduction of the efficiency by 1$\% $ has been
   simulated. No significant change in the signals at $SQ=0$ has been
   observed. Since R is a ratio with respect to the Monte Carlo simulation
   (including the detector effects) and the deviation which has been
   observed between data and MC below 10\%, it can be expected that
   efficiency effects cancel to a large extent.

\item[(d)]  
To investigate whether the good agreement between
data and Monte
   Carlo in quark jets is only due to the larger
   particle momenta, in a test-run
   only particles with momenta less than 30~GeV/c have been accepted in jet1.
   The agreement with the Monte Carlo 
   remains.

\item[(e)]
    The estimations leading to (11) and (12) assume that quark jets, also at
   the lower energies of jet3, do not exhibit any excess of neutral leading 
   systems. This is further tested by measuring the excess in sample3
   which exhibits at $\Delta y = 1.5$ an admixture of only 20\% gluon 
   jets. As expected, the signal is reduced, and is even 
   negative with large error: $R^{'}_{3}$(1.5) = -0.02 $\pm$ 0.02.
   Adopting the same procedure as in Section 4.3 by using matrix inversion
   with the measured values $R_i^{'}$(1.5), i= 1,2,3 for the
   3 selected samples as input, the resulting excess
   for {\em pure} quark and gluon jets could be estimated:
   $R_q^{'}$(1.5) = 0.00 $\pm$ 0.02,  $R_g^{'}$(1.5) = 0.11 $\pm$ 0.03 
   and $R_b^{'}$(1.5) = -0.06 $\pm$ 0.04.  These results do not show  
   any evidence that quark jets exhibit an excess of neutral leading systems  
   for the lower jet3 energies.

\item[(f)]
    At the generator level of JETSET and for pure gluon jets
    the effect of changing parameters within limits has been studied. 
    For example, different DELPHI tunings have been used,
    the DELPHI tuning \cite{detune} has been replaced by that of
    OPAL \cite{opx} and by the JETSET default\footnote{ All these studies have
    been done with BE correlations included.}, and the popcorn parameter
    has been varied. Some changes of
    $P(SQ,\Delta y = 1.5)$ at $SQ=0$ are revealed in
    gluon jets and to a lesser extent in pure quark jets. At a gap size of
    $\Delta y = 1.5$ a maximum variation of $R(\Delta y)$ of 
    0.027 is observed.
\end{itemize}

The systematic error from (f) amounting to $18\%$ is taken into 
account. This is a conservative estimate with a factor
0.68 of the maximum variation, corresponding to 1$\sigma$
of a Gauss distribution. 
The contributions from (a)-(e) are
negligible. The systematic errors for samples 1 and 2 are
estimated as follows:
\begin{itemize}
\item[1)]Sample 1

a) from the spread in Fig.4a at $\Delta y$ = 1.5:

$\Delta R^{'}$ = 0.017 

b) uncertainty in purity: 0.0026 (see Section 4.3)

c) uncertainty from (f): 0.018

\end{itemize}
By quadratically adding all 3 contributions a systematic error of
$0.025$ is obtained.

\begin{itemize}
\item[2)]Sample 2

a) from the spread in Fig.4b at $\Delta y$ = 1.5:

$\Delta R^{'}$ = 0.021

b) uncertainty in purity: 0.0028 (see Section 4.3)

c) uncertainty from (f): 0.019
\end{itemize}
By quadratically adding all 3 contributions a systematic error of
$0.028$ is obtained.

\section{Mass spectra}
Color octet neutralisation of the gluon field could produce a resonance
spectrum which differs from that of color triplet fragmentation
\cite{Ander} implemented in JETSET. In order to investigate in
which region of the mass spectrum the observed excess of the
leading neutral systems is located, the invariant mass ($M$)
distributions 
\begin{equation}
P(M) = \frac{N(M,SQ=0,\Delta y)}{N(\Delta y)}
\end{equation}
of the leading systems with total charge zero at 
$\Delta y = 1.5$ have been calculated and
compared with the mean values of models 1-3 for the two cases:
\begin{itemize}
\item[a.] $M$ is computed using only charged particles
with a momentum $p \geq$ 0.2 GeV/c and assuming pion mass. 
This distribution is shown in Fig.5a 
for gluon enriched jets and in Fig.5c for
quark jets. 
Two broad bumps can be observed in Figs.5a,c which are 
the result of a superposition of a rapidly decreasing two-particle
spectrum which dominates for $M \lesssim$ 1 GeV/c$^2$, with an increasing 
spectrum consisting of 4 and more particles.
 The latter dominates and peaks at $\sim$ 1.5 GeV/c$^2$.
The region below 0.8 GeV/c$^2$ consists only of two particles.
One peak around $M \sim$ 0.8 GeV/c$^2$
can be attributed to the $\rho$ resonance, another at
$M \le$ 0.5 GeV/c$^2$ to a reflection of $\eta$, $\eta^{\prime}$ and $\omega$.
The latter statement is corroborated by the fact that
in events with no neutrals, the peak at $M \le$ 0.5 GeV/c$^2$ vanishes.
Only in Fig.5a a third peak is indicated by the data points 
just below 1 GeV/c$^2$, in the region of the $f_0(980)$ resonance.
In this region 0.9 $\leq M \leq$ 1 GeV/c$^2$ the two particle contribution
amounts to about 70\%.
Figs.5b,d show the difference between the data and the Monte Carlo 
event sample. 
For gluon enriched jets the
distribution in Fig.5b exhibits possible evidence for a
mass enhancement\footnote{The experimental mass resolution is below 10 MeV/c$^2$,
the binwidth in Fig.5 is 50 MeV/c$^2$.}
in the region  0.9 $\leq M \leq$ 0.95 GeV/c$^2$, but no signal is seen in this
region for quark jets in Fig.5d. This peak is the remaining part of the original
peak in Fig.5a after the subtraction of a small and narrow 
but significant signal in the MC at 0.95-1.0 GeV/c$^2$.
Without emphasizing too much this remaining narrow peak 
in Fig.5b, it has to be noted
that it survived a quality cut (by accepting only jets with a polar angle
$\geq 50^{\circ}$)
well above 3$\sigma$, whereas all other
deviations from zero in isolated bins in Figs.5b,d were decreased. 
Whether this narrow signal in gluon enriched jets can be attributed to  
$f_0(980)$ production remains an open 
interesting question for future studies with increased statistics.
In the mass range below 
2 GeV/c$^{2}$ an overall excess is observed in Fig.5b.

\begin{figure}[tb]
\begin{center}
%\mbox{\epsfig{file=pet4n.eps,width=17cm}}
\mbox{\epsfig{file=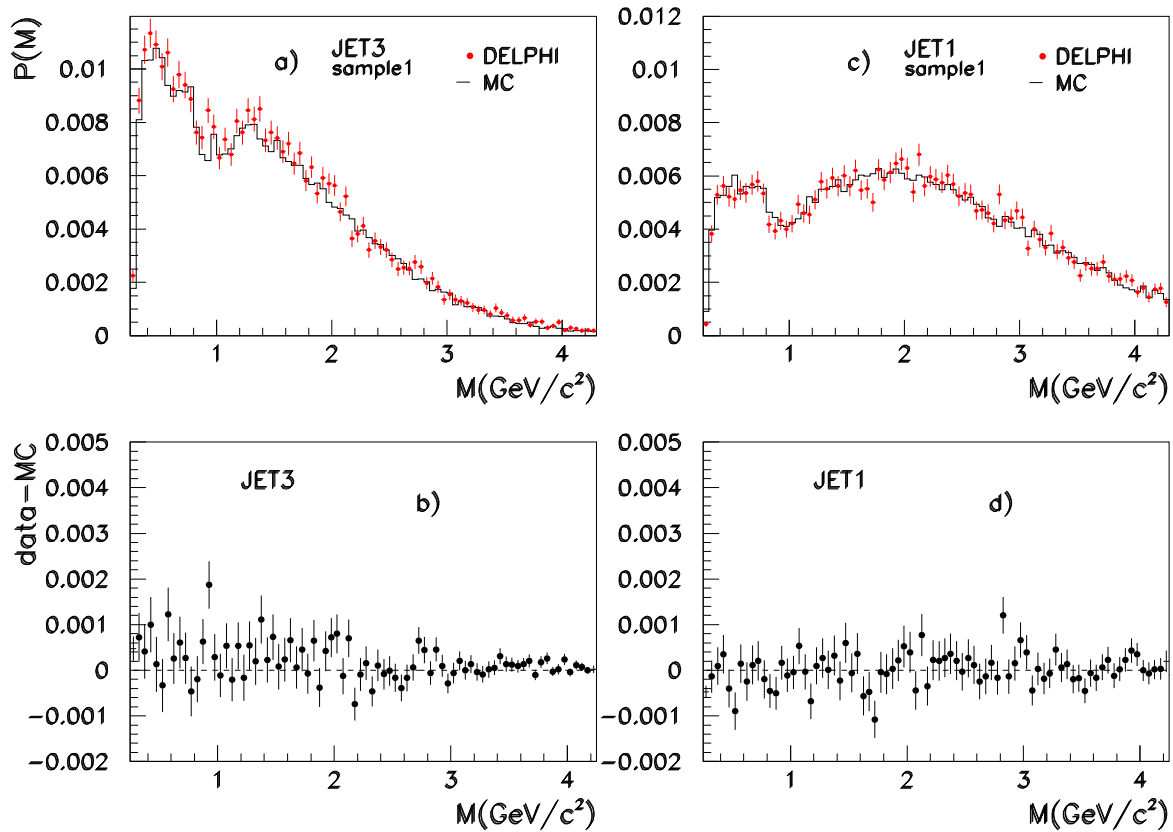,width=17cm}}
\caption[]{ a),c):
Invariant mass distribution $P(M)$ of the leading system ($SQ=0$)
(considering only charged particles) a)for gluon-enriched jets 
(sample1),
c) for quark jets. $P$ is defined in equ.(15).
The dots with error bars are the data, the histograms
are the mean values of the three Monte Carlos. b),d):
$P^{data}(M)-P^{MC}(M)$.}

\end{center}
\end{figure}

\item[b.] $M$ is computed for all charged particles
(with $p \geq$ 0.2 GeV/c and pion mass assumed) and neutrals 
(photons 
with $E \geq$ 0.5 GeV). 
Figs.6a,d show $P(M)$ for gluon enriched jets and Fig.6g for
quark jets.
Both gluon enriched samples exhibit an excess of data compared to the MC for
low invariant masses (below 2 GeV/c$^{2}$),
which is emphasized in Figs.6b,e where the mass
distributions of data and MC have been subtracted.
Separating the mass ranges below and above 2 GeV/c$^{2}$
in Fig.6c  for sample1, an excess of about 
5$\sigma$(stat) is observed
in the low mass range. 
Comparing Fig.6c with Fig.6f, one observes that the excess is increasing
according to increasing gluon purity, namely about a factor 2 between 
sample1  and sample2.
For quark jets, the corresponding distributions
(Fig.6g,h,i) do not exhibit any significant difference between
data and MC.\\
An important remark concerns all mass spectra in Fig.6. As stated in section 3,
all the comparisons to Monte Carlo event samples are done with full detector
simulation. Consequently, due to the loss of neutral particles, all 
spectra and in particular the excess spectra
in Figs.6b,6e are shifted by about 0.3 to 0.5 GeV/c$^2$ to lower mass values.
It has been verified, however, by a
special Monte Carlo study using the detector response matrix 
(not shown here) that, after the correction of all spectra for this shift,
the excess is still clearly concentrated in the low mass region.
%( $\leq$ 2.5 GeV/c$^2$ ) 
On the other hand, the spectra of leading systems consisting only of charged 
particles (Fig.5) are not 
affected by shifts.

\begin{figure}[!th]
\begin{center}
\mbox{\epsfig{file=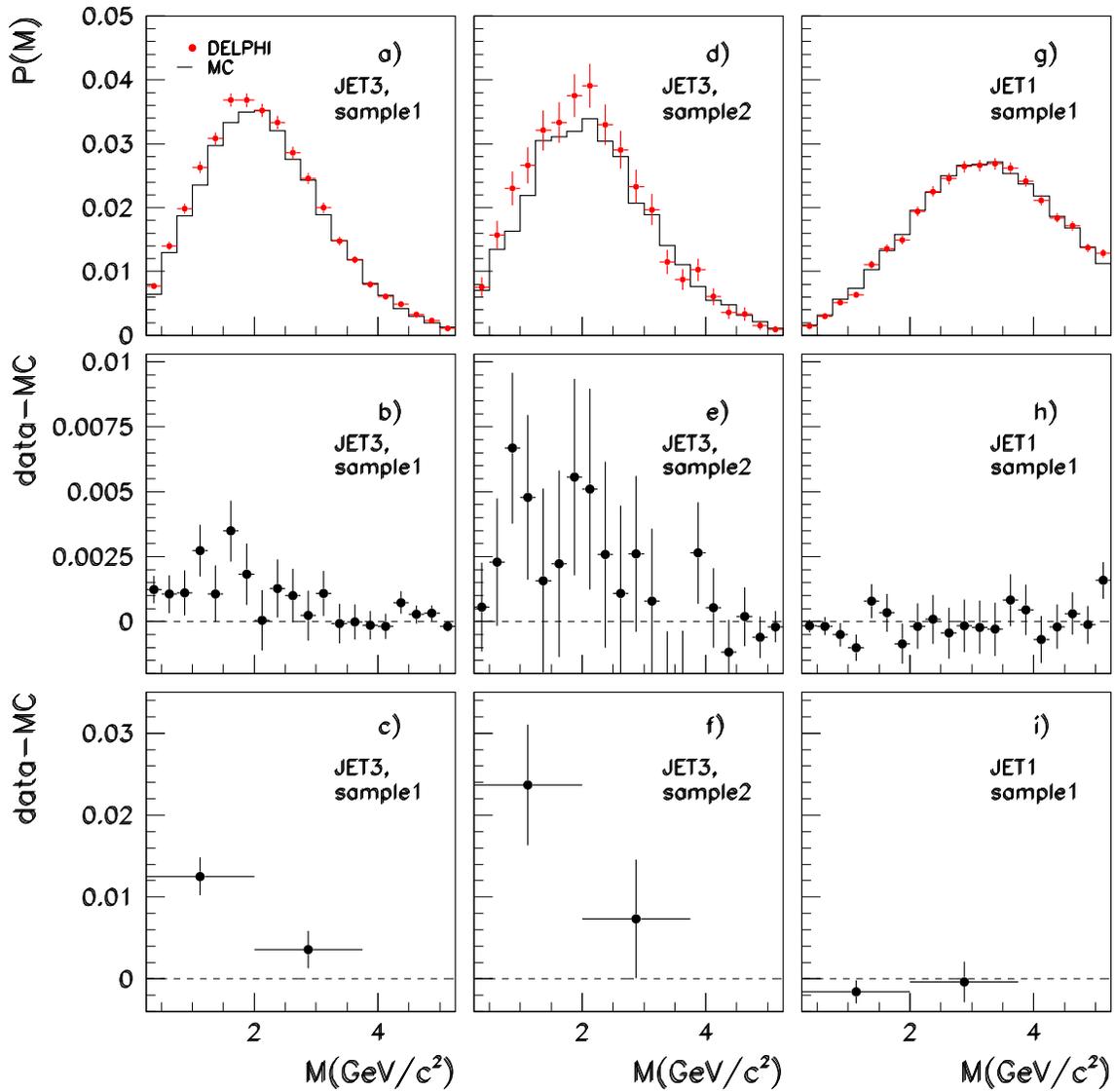,width=17cm} } \caption[]{ a),d):
Invariant mass distributions $P(M)$ 
of the leading system $(SQ=0)$ (with
charged and neutral particles) for gluon-enriched jets, samples 1 and 2 
respectively. g): same for quark jets (jet1, sample1).
The dots with error bars are the data, the histograms are
the mean values of the three Monte Carlos. The quantity $P(M)$ is defined 
in equ.(15). Second row b),e),h): 
$P^{data}(M)-P^{MC}(M)$;  Last row c),f),i):
$P^{data}(M)-P^{MC}(M)$ for  2 bins: (0.25-2 GeV/c$^2$) and
(2-3.75 GeV/c$^2$).} 
\end{center}
\end{figure}

\end{itemize}

The observations in Fig.5 confirm the first preliminary results 
presented in 2001
\cite{datong} for leading gluonic systems by considering charged
particles only. In 2002 the OPAL collaboration reported 
\cite{opalg} a 2$\sigma$ excess in the mass distribution 
of neutral leading systems,
consisting of charged and neutral particles, between
1 and 2.5 GeV/c$^2$ in gluon jets.

The observation  that the excess of neutral leading systems in
gluon jets is limited to the low mass region (Figs. 5 and 6)
supports arguments in favour of gluonic states. 
The existence of glueballs, i.e. bound states of two or more gluons,
is a prediction of QCD \cite{Fri}. 
The experimental results and their interpretations, however, are still
controversial \cite{Klem}. Theoretically there is general
agreement that the lightest glueball should be in the scalar
channel with J$^{PC}$ = 0$^{++}$. Quantitative results are derived from  
the QCD lattice calculations \cite{glust} 
which predict the lightest glueball to be around 1600 MeV/c$^2$, 
or from QCD sum rules \cite{qcds}, which predict also a possible 
gluonic state near
1 GeV/c$^2$. Alternatively, it could also be a very broad object 
\cite{An1,wope}.
%in the 
%mass region above 400 MeV/c$^2$ \cite{wope} or between 1200 and 1600
%MeV/c$^2$\cite{An1}. 
The gluonic state could mix with ordinary 0$^{++}$ states, like the
$f_0(980)$, $f_0(1370)$, $f_0(1500)$ or $f_0(1710)$. 
As an example there are scenarios, where the largest gluonic component
is included in the $f_0(1500)$ \cite{Am}, or alternatively in the
$f_0(1710)$ \cite{Sex}.
Recent discussions with various references
can be found in \cite{Klem,ochs3}.

\clearpage
\newpage

\section{Summary}
In the present study the leading systems defined by a rapidity
gap have been investigated for gluon and quark jets. The
statistics of 1994 and 1995 at $\sqrt{s}=91.2$~GeV obtained by the
DELPHI collaboration is used to select 3-jet events and to single
out quark jets (purity $\sim 90\%$) and gluon enriched jets (purity
$\sim 70\%$) by energy ordering (sample1). For the (enriched) gluon 
jets a higher rate of neutral leading systems than predicted by 
the Lund string model JETSET/ARIADNE 
(with and without Bose Einstein correlations)
is observed but no such enhancement is seen for the
quark jets. Various checks have been performed which suggest that
this effect is not a spurious one. An increase of the effect with
increasing gluon purity, obtained by a tagging procedure in a second
sample (sample2), is observed corroborating that it is indeed connected
with the  gluon jets.

The excess of neutral leading systems in pure gluon jets at a gap
size $\Delta y = 1.5$ has been measured to be about 10\%, 
with a significance of $3.6\sigma$. It is 
expected to be of the order of 0.5\% in pure
gluon jets  without any charge or gap selection.

The mass spectra of the neutral leading systems of gluon jets,
both with and without including neutral particles have been
studied. 
Mass spectra
which include charged and neutral particles, show clearly that
the excess mentioned above is concentrated at low
invariant masses ($\lesssim$ 2 GeV/c$^2$). The significance is enhanced
there and amounts to about 5$\sigma$(statistical) in sample1
and the excess is increased roughly proportionally to the 
gluon purity in sample2. 

The corresponding mass spectra
of leading systems in quark jets do not exhibit any
excess in the low  mass regions.

The observed  excess of neutral systems in gluon jets 
and its increase with the gap size and with the gluon purity
is in
agreement with expectations, if the hitherto unobserved but
predicted process of octet neutralisation of the gluon field 
takes place in nature.
Although color reconnection could in principle alternatively explain
the excess,
the specific mass concentration at low mass seems to
favor the first case and could be a signal of
gluonic states predicted by QCD.

%\input{acknow.tex}    % Achnowledgwement.
%         Modified on 04-06-1999 by dimartino
%-------------------------------------------------------------------
\subsection*{Acknowledgements}
\vskip 3 mm We thank W. Ochs for encouraging us to start the above
study and for discussions in the course of it and O. Klapp for
programming support. 
%and M.Siebel for valuable comments. 
We thank P. Minkowski, G. Rudolph and T. Sj\"{o}strand for valuable discussions and
clarifications. Technical support by G.Walzel is gratefully acknowledged.\\

We are greatly indebted to our technical collaborators, to the
members of the CERN-SL Division for the excellent performance of
the LEP collider, and to the funding agencies for their
support in building and operating the DELPHI detector.\\
We acknowledge in particular the support of \\
Austrian Federal Ministry of Education, Science and Culture,
GZ 616.364/2-III/2a/98, \\
FNRS--FWO, Flanders Institute to encourage scientific and technological
research in the industry (IWT), Belgium,  \\
FINEP, CNPq, CAPES, FUJB and FAPERJ, Brazil, \\
Czech Ministry of Industry and Trade, GA CR 202/99/1362,\\
Commission of the European Communities (DG XII), \\
Direction des Sciences de la Mati$\grave{\mbox{\rm e}}$re, CEA, France, \\
Bundesministerium f$\ddot{\mbox{\rm u}}$r Bildung, Wissenschaft, Forschung
und Technologie, Germany,\\
General Secretariat for Research and Technology, Greece, \\
National Science Foundation (NWO) and Foundation for Research on Matter (FOM),
The Netherlands, \\
Norwegian Research Council,  \\
State Committee for Scientific Research, Poland, SPUB-M/CERN/PO3/DZ296/2000,
SPUB-M/CERN/PO3/DZ297/2000, 2P03B 104 19 and 2P03B 69 23(2002-2004)\\
FCT - Funda\c{c}\~ao para a Ci\^encia e Tecnologia, Portugal, \\
Vedecka grantova agentura MS SR, Slovakia, Nr. 95/5195/134, \\
Ministry of Science and Technology of the Republic of Slovenia, \\
CICYT, Spain, AEN99-0950 and AEN99-0761,  \\
The Swedish Research Council,      \\
Particle Physics and Astronomy Research Council, UK, \\
Department of Energy, USA, DE-FG02-01ER41155, \\
EEC RTN contract HPRN-CT-00292-2002. \\

%=========================================================================%

%\input{acknow.tex}    % Achnowledgwement.
\end{document}